# Set, Reset, and Retention Times for Ionic and Filamentary Mem-Resistors

Blaise Mouttet

*Abstract*— A dynamic systems model has previously been proposed for mem-resistors based on a driven damped harmonic oscillator differential equation describing electron and ionic depletion widths in a thin semiconductor film. This paper derives equations for set, reset, and retention times based on the previously proposed model.

## I. INTRODUCTION

The analysis of [1] and [2] is based on several assumptions in accordance with small-signal ac behavior of ionic and filamentary memory resistance thin films. In order to calculate switching and retention times additional transient analysis is required. This paper includes such analysis.

## II. TRANSIENT ANALYSIS OF IONIC AND FILAMENTARY MEM-RESISTORS

As derived in [1] a dynamic equation may be formed for the time-dependent mean position $x_d(t)$ of ions at a boundary of a metal-semiconductor junction. The electron depletion approximation including a uniform ion donor density of $N_d(t)$ is assumed as illustrated in Fig. 1. The dynamic equation generally takes the form

$$\frac{d^2 x_d(t)}{dt^2} + \frac{1}{\tau_c}\frac{d x_d(t)}{dt} + \frac{(ze)^2 N_d x_{d0}}{2 m_{ion} \epsilon_r \epsilon_0}\left(1 - \frac{x_0(t)}{x_d(t)}\right) = \left(\frac{2av}{\tau_c}\right) exp\left(\frac{-W_a}{kT}\right) sinh\left(\frac{azeE_a(t)}{2kT}\right) \quad (1)$$

where $\tau_c$ is the average time between ion collisions, $z$ is the ion valence, $e$ is the unit charge, $N_d$ is the equilibrium ion donor density, $m_{ion}$ is the effective ion mass, $\varepsilon_r$ is the relative permittivity of the thin film material, $\varepsilon_0$ is the vacuum permittivity, $x_0(t)$ is the dynamic mean equilibrium of the ion positions,

$a$ is the jump distance of the ions between hopping, $v$ is a frequency factor, $W_a$ is the energy barrier, $k$ is Boltzmann's constant, $T$ is absolute temperature, and $E_a(t)$ is the electric field in the depletion region. This dynamic equation may be simplified under the following assumptions:

a) The mean equilibrium is time-independent $(x_0(t)=x_0=x_{d0}/2)$,

b) the variation $\Delta x_d(t)$ between $x_d(t)$ and $x_0$ is small compared to $x_0$,

$$\Delta x_d(t) = x_d(t) - x_0 \ll x_0 \quad (2)$$

Under these conditions (1) may be rewritten in terms of the mean ion position variation $\Delta x_d(t)$ as:

$$\frac{d^2 \Delta x_d(t)}{dt^2} + \frac{1}{\tau_c}\frac{d \Delta x_d(t)}{dt} + \frac{(ze)^2 N_d}{m_{ion} \epsilon_r \epsilon_0}\Delta x_d(t) = \left(\frac{2av}{\tau_c}\right) exp\left(\frac{-W_a}{kT}\right) sinh\left(\frac{azeE_a(t)}{2kT}\right) \quad (3)$$

where the mean equilibrium position $x_0$ of the ions is set equal to the electron depletion width $x_{d0}$ divided by 2.

*a) Retention time*

In order to approximate the retention time for a particular ion distribution state which differs from the equilibrium condition (i.e. $x_d(t) \neq x_0$) it may be assumed that the electric field $E_a(t)$ is near zero.

$$\frac{d^2 \Delta x_d(t)}{dt^2} + \frac{1}{\tau_c}\frac{d \Delta x_d(t)}{dt} + \frac{(ze)^2 N_d}{m_{ion} \epsilon_r \epsilon_0}\Delta x_d(t) = 0 \quad (4)$$

It is straight-forward to solve this differential equation producing

$$\Delta x_d(t) = c_1 \exp\left[\left(-\frac{1}{2\tau_c} + \sqrt{\frac{1}{4\tau_c^2} - \frac{(ze)^2 N_d}{m_{ion}\epsilon_r\epsilon_0}}\right)t\right]$$
$$+ c_2 \exp\left[\left(-\frac{1}{2\tau_c} - \sqrt{\frac{1}{4\tau_c^2} - \frac{(ze)^2 N_d}{m_{ion}\epsilon_r\epsilon_0}}\right)t\right] \quad (5)$$

where $c_1$ and $c_2$ are constants determined by the initial condition of the ion distribution. Assume a distribution such that at the initial time the mean ion deviation from equilibrium is $\Delta x_{d1}$ and the mean ion velocity is zero.

$$\Delta x_d(0) = \Delta x_{d1}$$
$$d[\Delta x_d(0)]/dt = 0 \quad (6)$$

Given this initial condition it is straight-forward to solve for $c_1$ and $c_2$ producing:

$$c_1 = \frac{\Delta x_{d1}}{2}\left[1 + \left(1 - \frac{4(ze)^2 N_d \tau_c^2}{m_{ion}\epsilon_r\epsilon_0}\right)^{-\frac{1}{2}}\right]$$
$$c_2 = \frac{\Delta x_{d1}}{2}\left[1 - \left(1 - \frac{4(ze)^2 N_d \tau_c^2}{m_{ion}\epsilon_r\epsilon_0}\right)^{-\frac{1}{2}}\right] \quad (7)$$

The retention time can be approximated from the coefficient of the first term exponential of (5) since the second term decays more quickly. For example the amount of time required for the first term exponential of (5) to reach 5% ($e^{-3}$) of the initial deviation is

$$t_{ret} \approx \frac{3}{\left(\frac{1}{2\tau_c} - \sqrt{\frac{1}{4\tau_c^2} - \frac{(ze)^2 N_d}{m_{ion}\epsilon_r\epsilon_0}}\right)} \quad (8)$$

It is possible to rewrite (5), (7), (8) in terms of ionic mobility $\mu_{ion}$ which may be more useful in some circumstances.

$$\mu_{ion} = (ze)\tau_c/m_{ion} \quad (9)$$

Plugging (9) into (5), (7), and (8) produces:

$$\Delta x_d(t) = c_1 \exp\left[\frac{ze}{m_{ion}}\left(-\frac{1}{2\mu_{ion}} + \sqrt{\frac{1}{4\mu_{ion}^2} - \frac{m_{ion}}{Ax_{d1}\epsilon_r\epsilon_0}}\right)t\right]$$
$$+ c_2 \exp\left[\frac{ze}{m_{ion}}\left(-\frac{1}{2\mu_{ion}} - \sqrt{\frac{1}{4\mu_{ion}^2} - \frac{m_{ion}}{Ax_{d1}\epsilon_r\epsilon_0}}\right)t\right] \quad (10)$$

$$c_1 = \frac{x_{d1}}{2}\left[1 + \left(1 - \frac{4(\mu_{ion})^2 m_{ion} N_d}{\epsilon_r\epsilon_0}\right)^{-\frac{1}{2}}\right]$$
$$c_2 = \frac{x_{d1}}{2}\left[1 - \left(1 - \frac{4(\mu_{ion})^2 m_{ion} N_d}{\epsilon_r\epsilon_0}\right)^{-\frac{1}{2}}\right] \quad (11)$$

$$t_{ret} \approx \frac{3}{\frac{ze}{m_{ion}}\left(\frac{1}{2\mu_{ion}} - \sqrt{\frac{1}{4\mu_{ion}^2} - \frac{m_{ion}N_d}{\epsilon_r\epsilon_0}}\right)} \quad (12)$$

It is evident from (12) that the retention time can be enhanced by a higher permittivity thin film. It is also notable from (12) that a lower dopant density of ions or oxygen vacancies ($N_d$) and a lower effective ion/vacancy mass will improve the retention time.

*b) Set and Reset time*

In order to calculate the amount of time required to set the ionic distribution to a state other than equilibrium (i.e. $x_d(t) \neq x_0$) or to reset the state back to equilibrium (i.e. $x_d(t) = x_0$) an external voltage needs to be considered. The relation between the electric field $E_a(t)$ and the applied voltage $V_a(t)$ in the depletion region may be approximated by

$$E_a(t) \approx -\frac{V_a(t)}{x_{d0}(t)} \quad (13)$$

As evident from Fig. 1 in which the mean ion position is half the depletion width the deviation of the depletion width is twice the deviation of the mean ion position so that

$$x_{d0}(t) = x_{d0} + 2\Delta x_d(t) \quad (14)$$

Combining (3), (13), (14) results in

$$\frac{d^2\Delta x_d(t)}{dt^2} + \frac{1}{\tau_c}\frac{d\Delta x_d(t)}{dt} + \frac{(ze)^2 N_d}{m_{ion}\epsilon_r\epsilon_0}\Delta x_d(t)$$
$$= -\left(\frac{2a\nu}{\tau_c}\right)\exp\left(\frac{-W_a}{kT}\right)\sinh\left(\frac{azeV_a(t)}{2kT(x_{d0}+2\Delta x_d(t))}\right) \quad (15)$$

Using the first two terms of a Taylor approximation around $\Delta x_d(t)$ for the forcing function produces

$$\frac{d^2\Delta x_d(t)}{dt^2} + \frac{1}{\tau_c}\frac{d\Delta x_d(t)}{dt} + \frac{(ze)^2 N_d}{m_{ion}\epsilon_r\epsilon_0}\Delta x_d(t) \approx$$
$$-\left(\frac{2a\nu}{\tau_c}\right)exp\left(\frac{-W_a}{kT}\right)[sinh\left(\frac{azeV_a(t)}{2kTx_{d0}}\right) - \quad (16)$$
$$cosh\left(\frac{azeV_a(t)}{kTx_{d0}}\right)\frac{azeV_a(t)}{kTx_{d0}^2}2\Delta x_d(t)]$$

The second term of the Taylor expansion is a linear function of $\Delta x_d$ and can be included as an additional factor affecting the resonant frequency $\omega_0$ of the harmonic differential equation.

$$\omega_0^2 = \frac{(ze)^2 N_d}{m_{ion}\epsilon_r\epsilon_0} - \left(\frac{4zea^2\nu}{x_{d0}^2\tau_c kT}\right)exp\left(\frac{-W_a}{kT}\right)cosh\left(\frac{azeV_a(t)}{kTx_{d0}}\right)V_a(t) \quad (17)$$

It is straight-forward to solve (16) in the same manner as the case used for the retention time except in this case the homogeneous solution will be dependent on the applied voltage via the resonant frequency term and there will be an additional particular solution term dependent on the sinh term. The reset times can be estimated in the same manner as the retention time using the coefficient of the more slowly decaying exponential of the homogeneous solution of (16). For example the amount of time required to reach 5% ($e^{-3}$) of the original deviation may be estimated as

$$t_{reset} \approx \frac{3}{\left(\frac{1}{2\tau_c} - \sqrt{\frac{1}{4\tau_c^2} - \frac{(ze)^2 N_d}{m_{ion}\epsilon_r\epsilon_0} - \left(\frac{4zea^2\nu}{x_{d0}^2\tau_c kT}\right)exp\left(\frac{-W_a}{kT}\right)cosh\left(\frac{azeV_a}{kTx_{d0}}\right)|V_a|}\right)} \quad (18)$$

wherein the ionic restoring force acts in the same direction as the reset voltage. For sufficiently high voltage such that the square root term approaches zero the set time would approach twice the average time between ion collisions and would be expected to be very small.

A similar analysis may be performed for a setting voltage in which case the ionic restoring force acts in the opposite direction as the set voltage. In this case the set time $t_{set}$ may be estimated as the time it takes to reach 95% ($1-e^{-3}$) of the desired mean ion position.

$$t_{reset} \approx \frac{3}{\left(\frac{1}{2\tau_c} - \sqrt{\frac{1}{4\tau_c^2} + \frac{(ze)^2 N_d}{m_{ion}\epsilon_r\epsilon_0} - \left(\frac{4zea^2\nu}{x_{d0}^2\tau_c kT}\right)exp\left(\frac{-W_a}{kT}\right)cosh\left(\frac{azeV_a}{kTx_{d0}}\right)|V_a|}\right)} \quad (19)$$

For the case of filamentary resistance mem-resistors as illustrated in Fig.2 the switching time is determined by the rate at which the ionic depletion region is created or destroyed. Increasing the electron depletion width will increase the built-in voltage and mobile ions will be drawn away from the gap between the filament tip and the electrode resulting in a creation of an ionic depletion region and a decrease in the gap distance. Conversely decreasing the electron depletion width will decrease the built-in voltage and mobile ions will be drawn toward the filament-electrode gap resulting in the destruction of the ionic depletion region. Because of this phenomenon the reset time for the electron depletion region and Schottky barrier will correspond to the set time for the tunneling barrier of the filament/electrode gap while the reset time for the tunneling barrier of the filament/electrode gap will correspond to the set time of the electron depletion region and Schottky barrier.

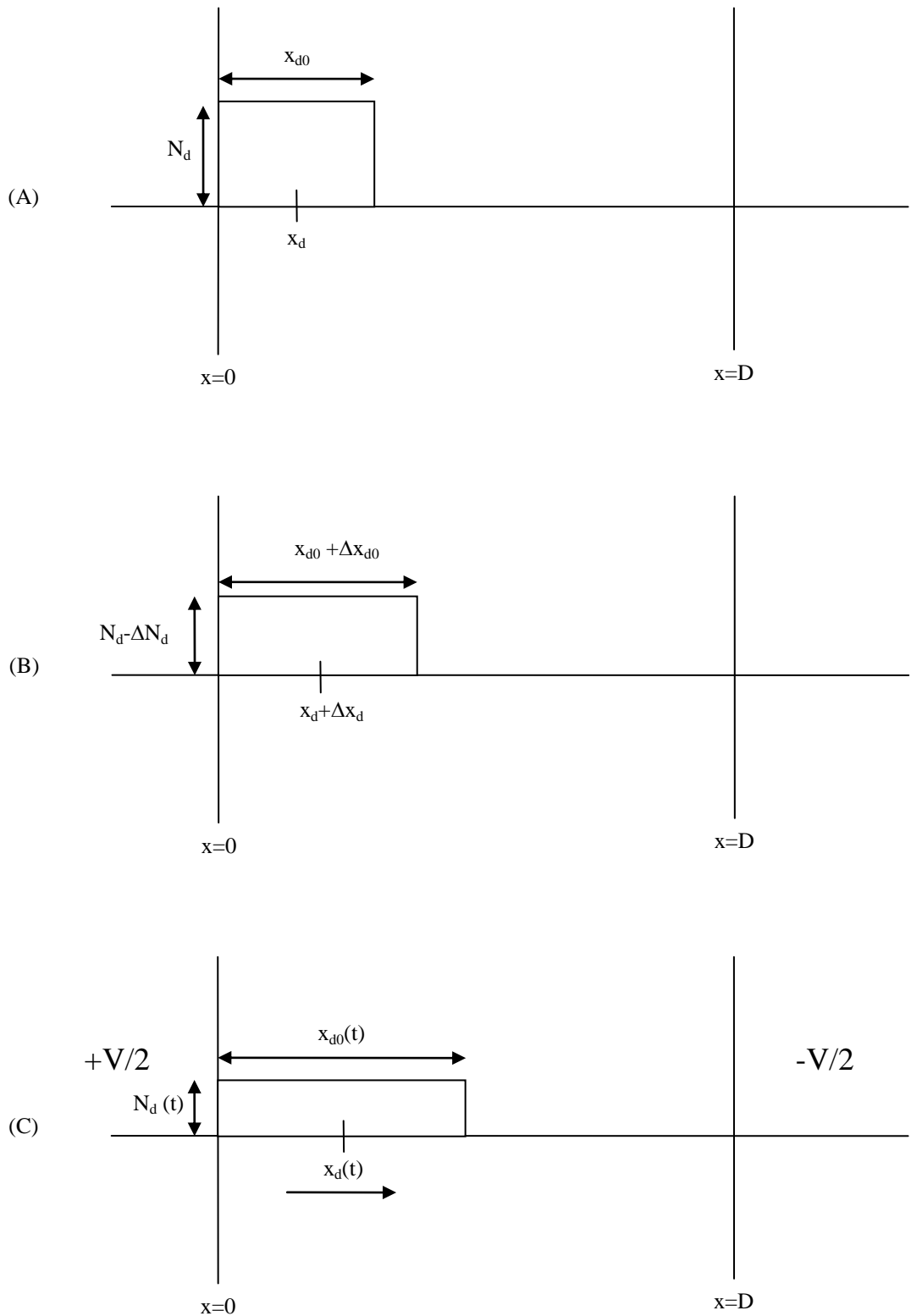

Fig. 1
(A) Uniform positive ion distribution at initial time with zero applied voltage. The mean of the distribution is located at $x_d$.
(B) At later time ions may spread out due to electrostatic repulsion but charge is conserved. $(N_d-\Delta N_d)(x_{d0}+\Delta x_{d0})=N_d x_{d0}$.
(C) A voltage is applied moving the mean of the ions toward the right. While the distribution $N_d(t)$ may change charge is conserved.

June 25, 2011 (ver 2)

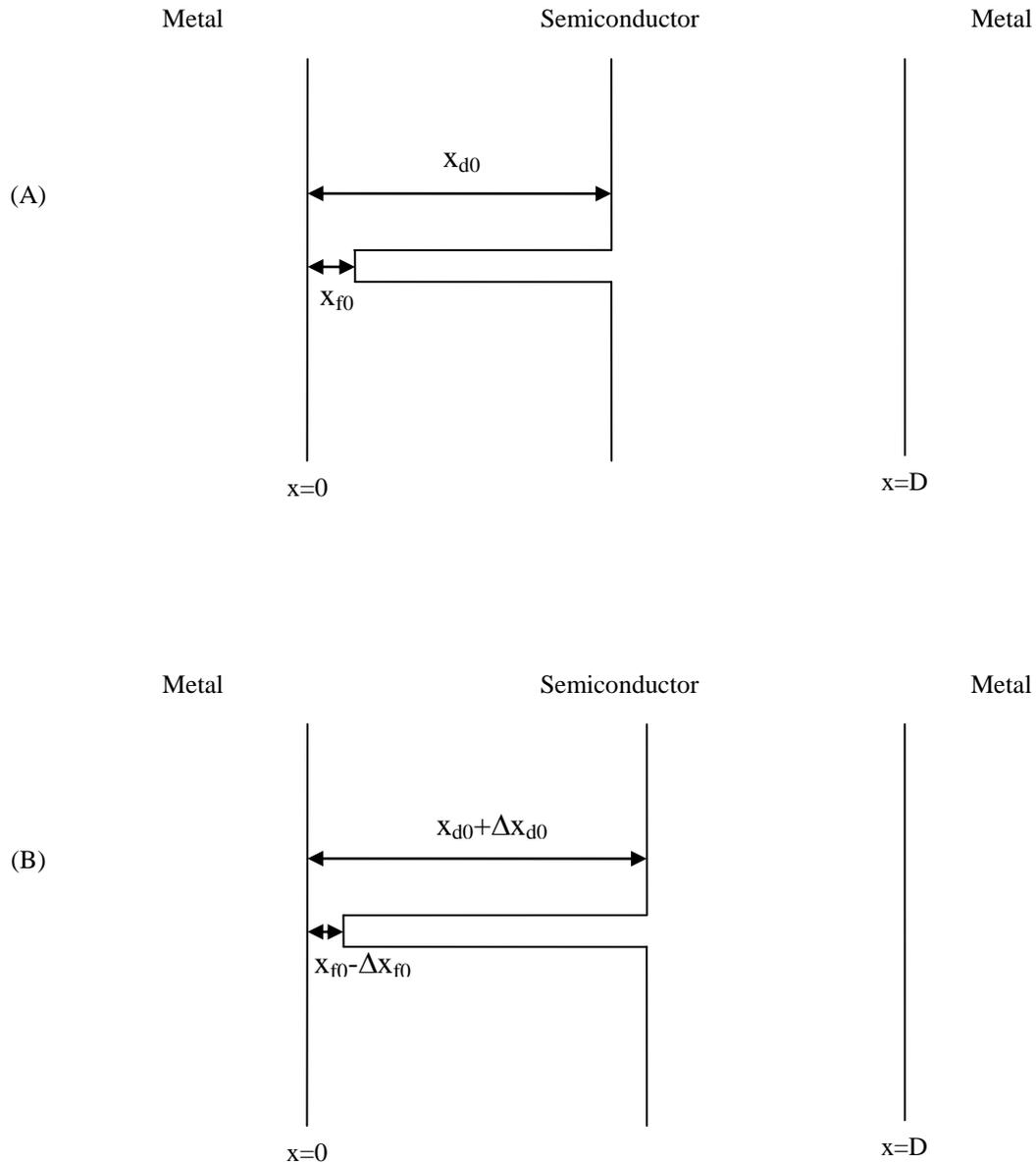

Fig. 2
(A) Illustration of a dynamic electron depletion width $x_{d0}(t)$ surrounding a filament and an ionic depletion width $x_f(t)$ between the filament tip and electrode in a metal-semiconductor-metal cell.
(B) Illustration of the coupled variation of electron and ionic depletion widths occurring in opposite directions.

June 25, 2011 (ver 2)